\documentclass[conference]{IEEEtran}
\usepackage{amsmath,amssymb,amsfonts}
\usepackage{algorithmic}
\usepackage{graphicx}
\usepackage{subcaption}
\usepackage{textcomp}
\usepackage{dblfloatfix}
\usepackage[subrefformat=parens,labelformat=parens]{subcaption}
\usepackage{xcolor}
\usepackage{cite}
 \usepackage{fontsize} 
\usepackage{subcaption}
\def\BibTeX{{\rm B\kern-.05em{\sc i\kern-.025em b}\kern-.08em
    T\kern-.1667em\lower.7ex\hbox{E}\kern-.125emX}}
    
\begin{document}

\title{Self-supervised learning of a tailored Convolutional Auto Encoder for histopathological prostate grading.}

\author{\IEEEauthorblockN{Zahra Tabatabaei\IEEEauthorrefmark{1}\IEEEauthorrefmark{4},
Adrián Colomer\IEEEauthorrefmark{1}\IEEEauthorrefmark{2},
Kjersti Engan\IEEEauthorrefmark{3}, Javier Oliver\IEEEauthorrefmark{4},  
Valery Naranjo\IEEEauthorrefmark{1}}

\IEEEauthorblockA{\IEEEauthorrefmark{1}\shortstack{ Instituto Universitario de Investigación en Tecnología Centrada en el Ser Humano,\\ HUMAN-tech, Universitat Politècnica de València, Spain}}
\IEEEauthorblockA{\IEEEauthorrefmark{2} ValgrAI – Valencian Graduate School and Research Network for Artificial Intelligence}
\IEEEauthorblockA{\IEEEauthorrefmark{3} Department of Electrical Engineering and Computer Science, University of Stavanger, Norway}
\IEEEauthorblockA{\IEEEauthorrefmark{4} Department of Artificial Intelligence, Tyris Tech S.L.,Valencia, Spain}}

\maketitle

\begin{abstract}
According to GLOBOCAN 2020, prostate cancer is the second most common cancer in men worldwide and the fourth most prevalent cancer overall. For pathologists, grading prostate cancer is challenging, especially when discriminating between Grade 3 (G3) and Grade 4 (G4). This paper proposes a Self-Supervised Learning (SSL) framework to classify prostate histopathological images when labeled images are scarce. In particular, a tailored Convolutional Auto Encoder (CAE) is trained to reconstruct 128×128×3 patches of prostate cancer Whole Slide Images (WSIs) as a pretext task. The downstream task of the proposed SSL paradigm is the automatic grading of histopathological patches of prostate cancer. The presented framework reports promising results on the validation set, obtaining an overall accuracy of 83\% and on the test set, achieving an overall accuracy value of 76\% with F1-score of 77\% in G4.
% F1-score of 71\%. 
% This model outperforms previous well-known arthitecures in discriminating G3 and G4 in prostate histopathological images with 77\% of F1S. 
\end{abstract}

% \begin{keywords}
% Classification, Convolutional Auto Encoder (CAE), Histopathological images, Prostate cancer, Whole Slide Images (WSIs).
% \end{keywords}
\section{Introduction}
Prostate cancer is the fifth most deadly disease, according to GLOBOCAN 2020, having an estimation of 375,304 cases. Prostate cancer tends to spread slowly, and in some cases, men who died of other diseases also had prostate cancer that had never impacted them in life \cite{marar2022outcomes}. In the traditional cancer diagnosis, if the therapist confirms prostate cancer based on blood tests, small portions of the cancerous tissue are extracted. These tissues need to pass some processing, and then they should be stained. Hematoxylin and Eosin (H\&E) are one of the most widely used materials in staining tissues, and it provides pink and purple slides for pathologists\cite{kanwal2022quantifying}. Pathologists grade the tissues by examining them under a microscope to analyze how the glands, nuclei, and lumen are organized. Prostate cancer has four classes Non-Cancerous (NC), Grade 3 (G3), Grade 4 (G4), and Grade 5 (G5). The presence of one or more Gleason patterns in the tissue regions is an interesting element in diagnosing the Gleason score. This score defines the proper treatment to apply; however, in some cases detecting the abnormalities is complicated, and it is not as easy as looking at the microscope \cite{going}. 
% Although the four grades of this malignancy have very slight distinctions and can often make grading challenging, each grade requires a different course of treatment. 
\begin{figure*}[b!]
\centerline{\includegraphics[width=0.66\textwidth]{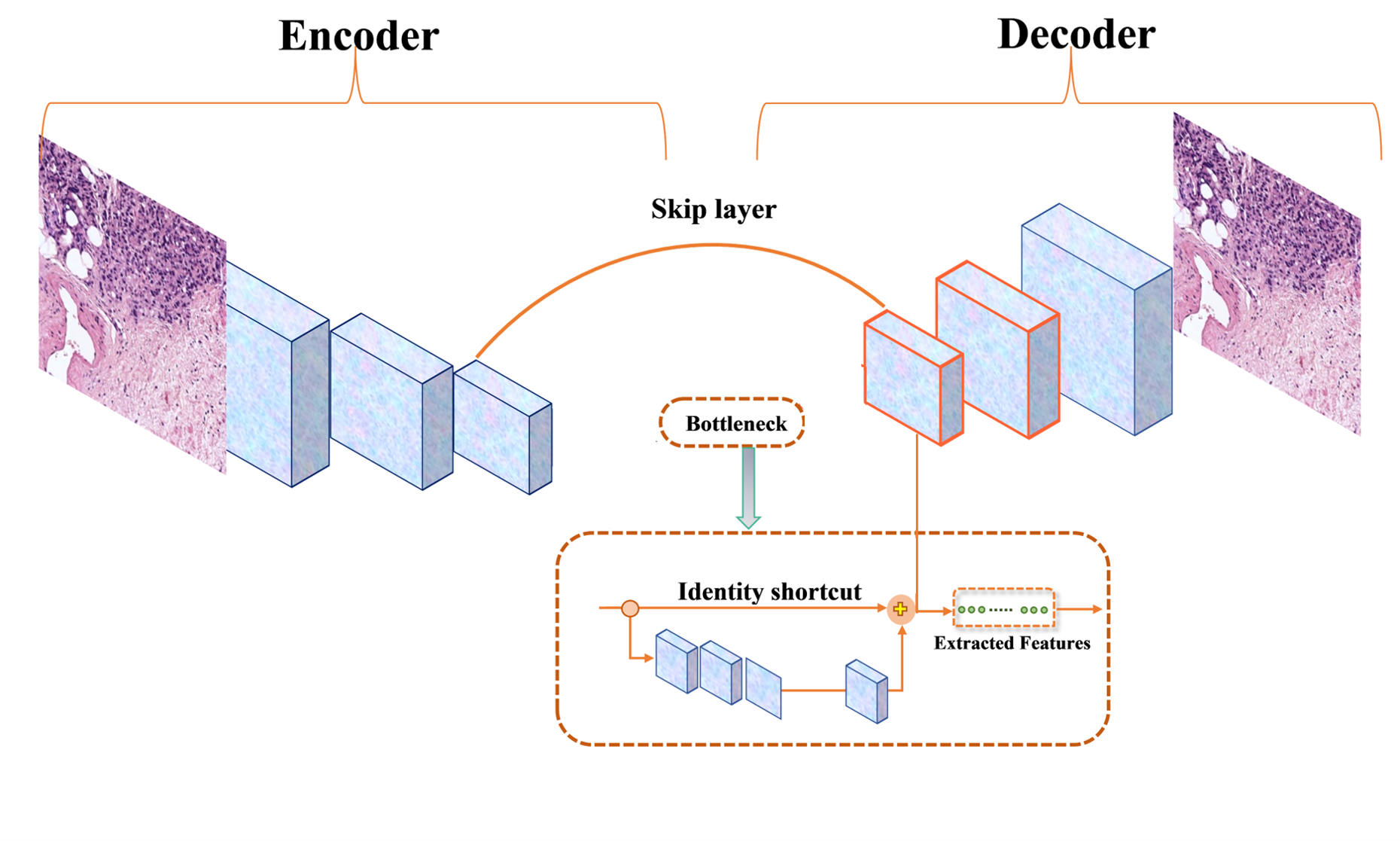}}
\caption{The structure of the proposed CAE as the pretext task. 
The intermediate layers consist of a skip layer to transfer information between corresponding encoder-decoder blocks \cite{9816325}. The bottleneck has four convolutional layers with stride = $1$ and kernel size = $(3,3)$.}
\label{fig:ThestructureoftheproposedCAE.}
\end{figure*}
To tackle these difficulties, the Computer-Aided Diagnosis system (CAD) provides a wide diversity of tools on scanned Whole Slide Images (WSIs) to increase the accuracy of a cancer diagnosis. These tools use Deep Learning (DL) and convolutional layers to extract features of the histopathological images, including color, texture, shape, size, etc. These features feed to fully connected layers to train the model that classifies the patches into different grades. Different cancer types have varying grades, and some of them only fall into the malignant and benign categories, necessitating a binary classification. To categorize the patches into more than two classes, multi-class classifiers are required.

The analysis of histopathological images has been done in previous literature using a variety of techniques. Spanhol, \cite{7727519} proposed a method to allow using the high-resolution histopathological images from the BreakHis data set without computational costs in the architecture. Liu, \cite{info13030107} proposed an improved Auto Encoder (AE) network to have an automated binary classification on the BreakHis data set. Silva-Rodr{\'\i}guez, \cite{going} proposed a patch-wise predictive model based on a Convolutional Neural Network (CNN) to grade the WSIs relating to the presence of cancerous patterns. Schmidt, \cite{schmidt2022efficient} presented cancer classification by coupling semi-supervised and multiple instances learning on prostate and breast cancer.
Kalapahar, \cite{kalapahar2020gleason} fed residual U-NET by SICAPv2 to grade the WSIs through semantic segmentation. Due to the similarities of some Gleason patterns in prostate histopathological tissues, multi-class classification is an open challenging task. Whereas both pathologists and DL techniques find it more challenging to grade prostate tissues into G3 and G4 than other grades, G3 and G4 need different medical treatments.

% The aforementioned works have in common that the label of input images are the output of the classifiers. The major objective is clearly stating that each patch belongs to which class. 
% In the aforementioned works, the major objective is labeling the input patches. This can assist pathologists in ensuring that their grading is accurate, and it is especially helpful for recently graduated pathologists with less experience. 
% In this work, in contrast with the cited above, we present a novel Self-Supervised Learning (SSL) strategy to classify prostate patches. 

In this work, we propose a novel Self-Supervised Learning (SSL) strategy to classify prostate patches into four grades. Our proposed SSL classifier, in contrast with the methods cited above, can mitigate this problem with an accurate delimitation of the different grades. More precisely, we train an SSL classifier containing a tailored CAE in its pretext task and a stack of fully connected layers as a classifier module for the downstream task in the place of the decoder in the CAE. Hence, the proposed classifier can detect very slight distinctions between grades, and it can provide a precise delineation of the complex patterns of G3 and G4. Compared to the prior patch-level techniques, the proposed SSL classifier performs comparably well at distinguishing G3 and G4. 
% The superiority of using our SSL classifier is that this model has been trained on histopathological images which makes it more reliable than a pre-trained model with a bunch of simple and non-medical images like animals or vehicles.

\section{Material}
SICAPv2 is the biggest public data set of prostate cancer containing pixel-wise annotations of the tumor regions following the Gleason score scale. This contains 182 prostate biopsies from 96 patients. A group of pathologists from Hospital Clínico of Valencia analyzed the H\&E stained images at 40$\times$ magnification \cite{9816325}. Table \ref{tab1:SICAPv2 data set description.} describes the number of WSIs and patches in SICAPv2 per grade.

One of the challenges in working with WSIs is their big size. Thus, they were re-sampled into 10x magnification and divided into patches of 512$\times$512$\times$3 pixels by a sliding window mechanism with 50\% of overlapping \cite{going}. In this paper, regarding constraints computation time, the images were resized to 128$\times$128$\times$3.

\begin{table}[ht]
\caption{SICAPv2 data set description.}
% \begin{adjustbox}{width=0.7\columnwidth}
\begin{center}
\begin{tabular}{|c|c|c|c|c|}
\hline
\textbf{Grades}&\textbf{NC} & \textbf{G3} & \textbf{G4} & \textbf{G5} \\
\cline{1-4} 
\hline\hline
WSIs & 37 & 60 & 69 & 16\\
\hline
Patches & 4417 & 1636 & 3622 & 655\\
\hline
\end{tabular}
\label{tab1:SICAPv2 data set description.}
\end{center}
% \end{adjustbox}
\end{table}
\section{Method}
\begin{figure*}[!ht]
\centerline{\includegraphics[width=0.66\textwidth]{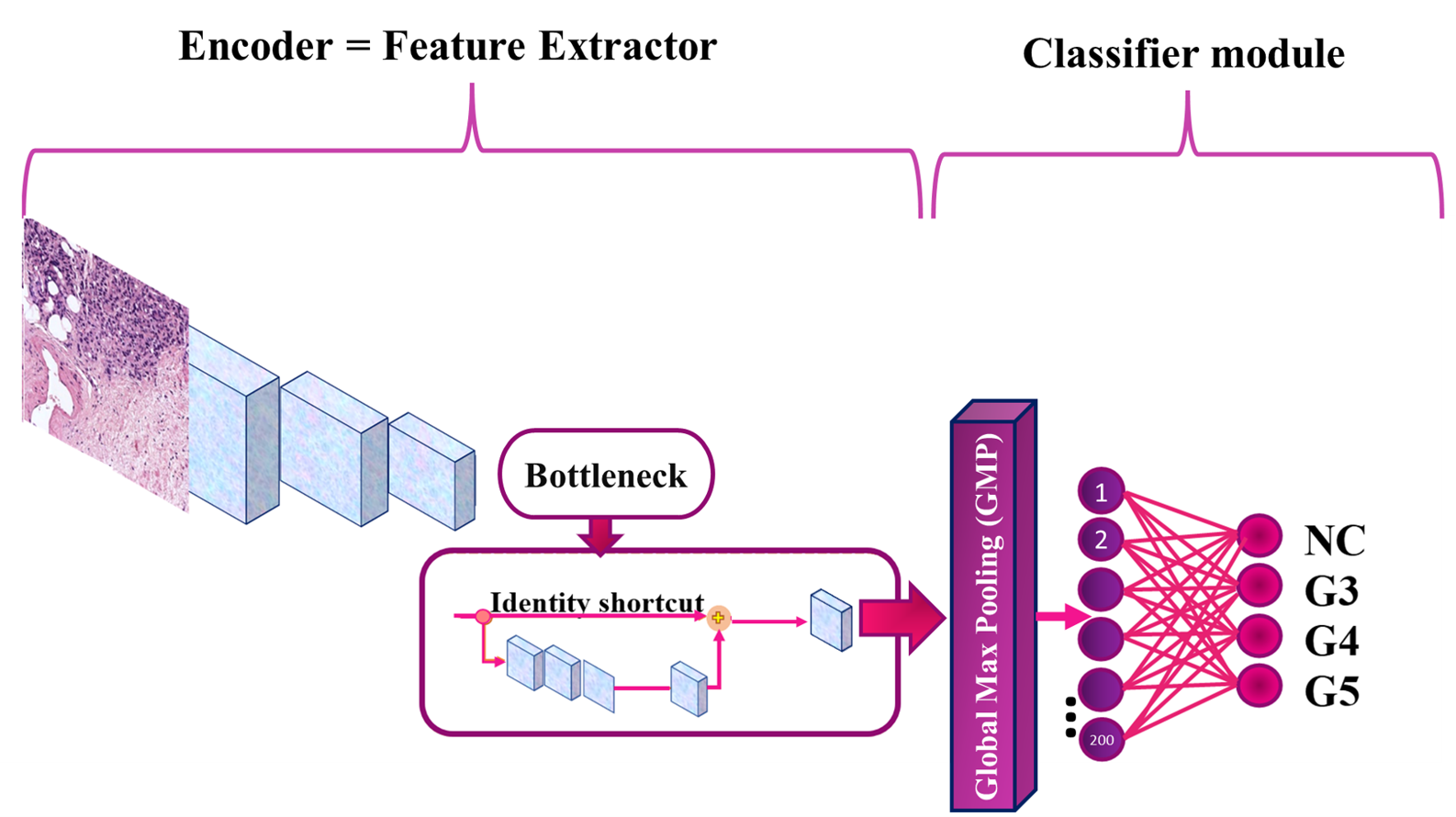}}
\caption{The structure of the proposed classifier on SICAPv2 as the downstream task (real task). The extracted features from the pretext task are fed into the downstream task, and the output layer presents one neuron per target grade with soft-max activation.}
\label{fig:The structure of the proposed classifier.}
\end{figure*}

%  presenting the methodological core of this work. It is an end-to-end patch-level classifier which is carried out by means of CAE in SSL manner.
The proposed end-to-end patch-level framework based on a tailored CAE trained in an SSL strategy is depicted in Fig.~\ref{fig:ThestructureoftheproposedCAE.} and Fig.~\ref{fig:The structure of the proposed classifier.}.

\subsection{Self Supervised Learning (SSL)}
%Supervised learning has pre-defined labels, however, unsupervised learning has just data without labels. 
Supervised DL has achieved great success in the last decade, but it depends on manual labeling and vulnerability. Learning with scarce labeled data is a fundamental problem in DL. Also, it is critical for histopathological image analysis because annotating medical images is time-consuming and expensive. The emergence of SSL tackled this problem.  
\begin{table*}[b]
\begin{center}
\caption{The obtained results of patch-level classification by our proposed classifier in the validation set of SICAPv2. The presented metrics are accuracy (ACC), F1-Score (F1S) computed per class and its average (F1-avg), and Cohen’s quadratic kappa (\textit{k}).}
\label{tab3:Results of SICAPv2 on validation set}
\centering
\begin{tabular}{lclclclclclclcl}\hline\hline
\textbf{Method}&\textbf{ACC}&\multicolumn{4}{c}{\textbf{F1S}}&\textbf{F1-avg}&\textbf{\textit{k}}\\
\cline{1-7} 
\hline
Ours &\textbf{0.83}  & \textbf{0.92} & \textbf{0.73} & \textbf{0.82} &\textbf{0.67} &\textbf{0.78}& \textbf{0.76} \\
VGG19 + FC \cite{going} & 0.72 &  0.88 & 0.66 & 0.60 & 0.52  & 0.66 & 0.73\\
VGG19 + GMP \cite{going} & 0.72 &  0.87 & 0.64 & 0.60 & 0.54  & 0.66 & 0.71\\
ResNet + FC  \cite{going} & 0.69 &  0.83 & 0.66 & 0.57 & 0.48  & 0.64 & 0.68\\
ResNet + GMP \cite{going}& 0.68 &  0.83 & 0.64 & 0.55 & 0.50  & 0.63 & 0.67\\

\hline\hline
\end{tabular}
\end{center}
\end{table*}
SSL has soared in performance on representation learning in the last several years. SSL is appealing because it allows for the pre-training usage of unlabeled domain-specific images in order to learn more appropriate representations \cite{azizi2021big}. It derives its labels from a co-occurring modality for the given data. To be able to understand the geometry of the input, SSL learns the representation of the data by observing different parts of it \cite{chen2020simple}. Generally, SSL includes a pretext and a downstream (real) task. In the pretext task, the visual representations are learned to have the obtained model weights for the downstream task. SSL leverages input data as supervision in pretext and benefits almost all types of downstream tasks. Then, the representations in the downstream task can be fine-tuned depends on the goal of the system with a few labeled images \cite{devlin2018bert, liu2021self}. In this work, the downstream task is classification with insufficient annotated patches.

\subsection{Pretext and downstream tasks for prostate cancer gradation}
% FIG illustrates an overview of the whole pipeline in the proposed classifier.
The intuition behind the proposed method is image reconstruction by a tailored Convolutional Auto Encoder (CAE) and classification by a multi-layer perceptron from a latent space as its pretext and downstream task, respectively. Pretext task has three benefits in our work including the classifier's initial weights are not random, the classifier is aware of the general characteristics of the WSIs before the first epoch, and it takes less time to categorize the images.
% One important aspect of SSL and fine-tuning using the weights of the pretext as initialization for the downstream supervised task is the choice of a proper unlabeled data set. 
These benefits are attained because, in contrast to other transfer learning techniques, our model is fed by histopathological images in pretext. 

% \subsubsection{Pretext: Convolutional Auto Encoder}
We proposed a tailored CAE in the pretext task to deal with the complexity of the histopathological images. As can be seen in Fig.~\ref{fig:ThestructureoftheproposedCAE.} an encoder's middle block is connected to the corresponding decoder's block through a skip connection. In order to extract more representative features, a residual block is added to the bottleneck. Notably, the tailored CAE compresses the input and reconstructs it using convolutional layers without a pooling layer. In the pretext stage, the decoder has thus been dropped after pre-training, and the encoder and bottleneck are set as the Feature Extractor (FE) for the classifier in the downstream task. In the downstream task, FE and the classifier module have been trained together to minimize the amount of categorical cross-entropy in the loss function. The architecture of the proposed classifier in Fig.~\ref{fig:The structure of the proposed classifier.} demonstrates how each image in the input is categorized into the corresponding label.

\section{Experiments and results}
We perform a series of experiments to train and validate the proposed classifier by splitting the data set into 80:20 for training and validating, respectively. In order to perform a comparison of the different classifiers described previously, the test set is completely isolated.

In particular, the tailored CAE was trained using Adam optimizer with a learning rate of $5\times 10^{-4}$ and a batch size of 16 images. Then, in the downstream task, weights of the first 29 levels of the encoder to the first 29 layers of the classifier were loaded. The bottleneck's output was fed into the classifier component, which has a Global Max Pooling 2D (GMP) layer and two dense layers $[200, 4]$ to obtain the probability distribution of four categories. In downstream, the classifier has was with a batch size of 8 images, a learning rate of $0.5$, and an SGD optimizer. 
\begin{figure*}[t]
    \centering
    \subfloat[\centering \label{fig:TSNE_our}]{{\includegraphics[width=.40\linewidth]{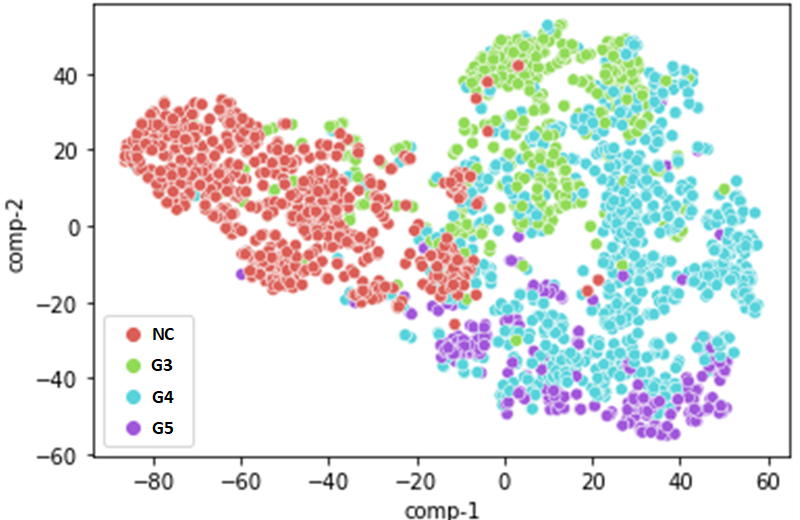} }}%
    \qquad
    \subfloat[\centering \label{fig:TSNE_arne}]{{\includegraphics[width=.40\linewidth]{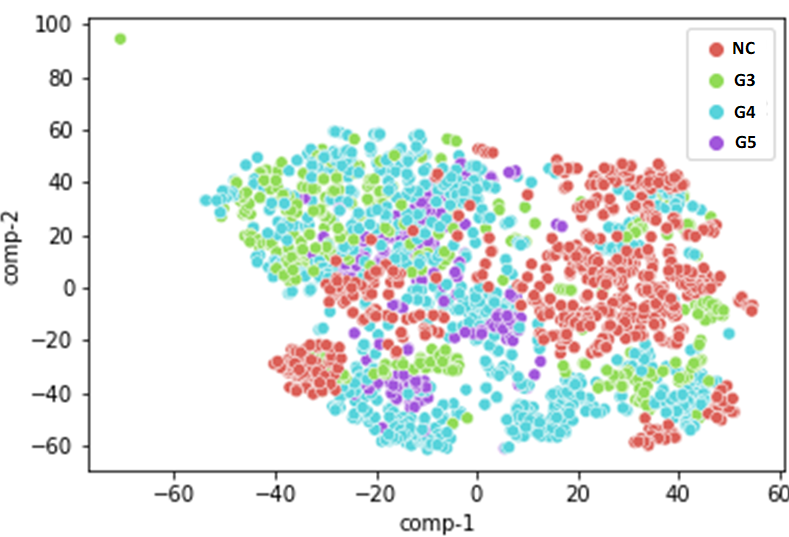} }}%
    \caption{TSNE projection of the extracted features of the test set in a 2D representation. Fig. 3(a) and Fig. 3(b) respectively represent the feature space as a result of the proposed method and Schmidt, Arne \emph{(2022)} \cite{schmidt2022efficient}.}
    \label{fig:TSNE}%
\end{figure*}

% The test set is completely isolated from the two other portion of data set. 
To establish a proper comparison with previous work, five indicators were selected, including accuracy (ACC) \cite{luque2019impact}, F1-score (F1) \cite{cuadros2016quality}, F1-avg, Kappa (\textit{k}), and Confusion Matrix (CM). Table \ref{tab3:Results of SICAPv2} and Table \ref{tab3:Results of SICAPv2 on validation set} summarize that our model reaches a promising performance on both validation and test set while it does not need any pre-trained model with a huge number of images. Comparing the \textit{k} values in the test set, which is reported in Table\ref{tab3:Results of SICAPv2} demonstrates how our suggested method is superior to recent works.
\begin{table}[h!]
\begin{center}
\caption{The obtained results of patch-level classification by our proposed classifier in the test set of SICAPv2. The presented metric is Cohen’s quadratic kappa (\textit{k}).}
\label{tab3:Results of SICAPv2}
\centering
\begin{tabular}{lclcl}\hline\hline
\textbf{Method}&\textbf{\textit{k}}\\
\cline{1-2} 
\hline
Ours & \textbf{0.80} \\
FS\textit{conv}+GMP \emph{(2020)} \cite{going} &0.77 \\
Nir et al. \emph{(2018)} \cite{nir2018automatic} & 0.61\\
Otalora et al. \emph{(2020)} \cite{otalora2020semi}& 0.55/0.59\\ 
Inter-Pathologists \emph{(2018)} \cite{arvaniti2018automated}& 0.65\\ 
Arviniti et al. \emph{(2018)} \cite{arvaniti2018automated} & 0.55/0.49\\ 
\hline\hline
\end{tabular}
\end{center}
\end{table}

\begin{figure}[h!]
     \centering
     \begin{subfigure}[b]{0.35\textwidth}
         \centering
         \includegraphics[scale = 0.47]{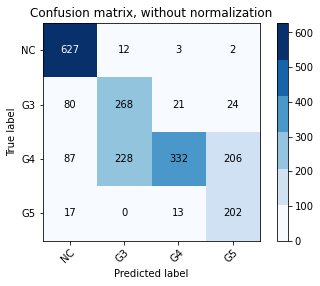}
         \caption{}
         \label{fig:y equals x}
     \end{subfigure}
     \\[\smallskipamount]
     \begin{subfigure}[b]{0.35\textwidth}
         \centering
         \includegraphics[scale = 0.47]{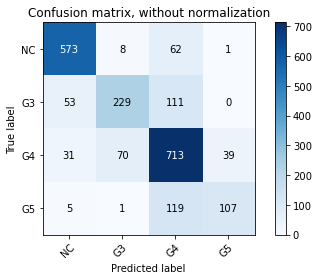}
         \caption{}
         \label{fig:three sin x}
     \end{subfigure}
     \caption{CMs of the grading by (a) proposed method in \cite{going} and (b) our proposed classifier.}\label{fig:CMs}
\end{figure}
In Table \ref{tab3:Results of SICAPv2 on validation set}, the performance of different models such as ResNet, VGG19, and FSConv are presented with different configurations of top models. We observe that our model can achieve competitive results with respect to other methods, such as VGG19, that had been pre-trained on a large number of non-histopathological images. Our SSL classifier measures \textit{k} better than more contemporary approaches, as shown in Table \ref{tab3:Results of SICAPv2 on validation set}, which supports the efficacy of the suggested SSL algorithm on a small training data set. The F1S for each grade in the validation set demonstrates that the proposed method is more effective in distinguishing between the four grades, particularly in the challenging task of differentiating between G3 and G4. This suggests that the model is more robust in its ability to classify the grades accurately.
% \begin{figure}[t]
%      \centering
%      \begin{subfigure}[b]{0.28\textwidth}
%          \centering
%          \includegraphics[scale = 0.47]{images/TSNE_rewise_CAE_paint.png}
%          \caption{}
%          \label{fig:TSNE_our}
%      \end{subfigure}
%      \\[\smallskipamount]
%      \begin{subfigure}[b]{0.28\textwidth}
%          \centering
%          \includegraphics[scale = 0.47]{images/TSNE_rewise_arne_paint.png}
%          \caption{}
%          \label{fig:TSNE_arne}
%      \end{subfigure}
%      \caption{CMs of the grading by (a) proposed method in \cite{going} and (b) our proposed classifier.}\label{fig:TSNE}
% \end{figure}

In order to analyze the correlation of the extracted features for each class, T-distributed Stochastic Neighbor Embedding (TSNE) technique was applied to them, Fig.~\ref{fig:TSNE}. Fig. 3(a) shows a 2D TSNE plot, representing features of the four classes under this study. The plot shows a discriminative behavior of the proposed framework, allowing a promising gradation of prostate histopathological patches with a F1-avg value of 72\% on the test set. As can be seen in Fig. 3(a) compared to Fig3(b) \footnote {Fig. 3(b) was plotted based on the features shared by Schmidt, Arne \cite{schmidt2022efficient}}, our proposed model can reduce the inter-distance and can increase intra-distance between features of each class much better than the recent work \cite{schmidt2022efficient}. In Fig. 3(a), the extracted features of G4 are more spread in the feature space, while it brings F1S equal to 77\% on the test set, which is higher than compared methods. 

Most notably, from a pathologist's point of view, G3 and G4 are slightly different. However, G3 must take medication, and G4 must have surgery. This emphasizes the significance of accurately grading G3 and G4. In Fig. \ref{fig:CMs} CMs show that the model classifies the images into four classes more accurately and it can surpass and outperform the reported CM in \cite{going}, especially in grading G3 and G4.

\section{Conclusions}
% We have presented an end-to-end DL structure that is able to classify the data set very accurately while does not require a pre-trained model on a pool of images.
% The highlighted power points in our method are, first, our method starts training the classifier not from scratch with random weights. Second, the initial weights in training the proposed CAE classifier are not from irrelative and non-medical/ histopathological images. The superiority of using our own SSL classifier is that this model can be more trustworthy than using a pre-trained model which is first trained to classify a bunch of simple images like animals or vehicles. 
% We have presented a novel end-to-end SSL model that is able to classify the data set with accuracy of.... while does not require a pre-trained model on a pool of images. The proposed SSL method enables to learn the first meaningful features of WSIs by training a tailored CAE to reconstruct the images in the pretext task. 
In this paper, we have presented a novel end-to-end DL structure that is able to classify a multi-class data set without the need for a pre-trained model on a pool of images. The proposed SSL method enables learning the first meaningful features of WSIs by training a tailored CAE to reconstruct the images in the pretext task. In our downstream task, we can get a classification accuracy of 76\% in the test set, surpassing competing methods in the same and other data sets and indicating the value of our method. The highlighted power point in our method is an accurate delimitation of the complex patterns of G3 and G4 in prostate tissues resulting in an F1S value of 77\%, which outperforms the previous works.

% first, our method does not start training the classifier from scratch with random weights. Second, the initial weights in training the proposed classifier are not from non-medical/histopathological images.  

% \section{Compliance with ethical standards}
% This research study was conducted retrospectively using human subject data made available in open access by (\url{https://data.mendeley.com/datasets/9xxm58dvs3/1)}. Ethical approval was not required as confirmed by the license attached with the open access data.
% \label{sec:ethics}

% \shortstack{https:\\//data.mendeley.com/datasets/9xxm58dvs3/1}
\section{Acknowledgments}
\label{sec:acknowledgments}

In this study, Zahra Tabatabaei is funded by European Union’s Horizon 2020 research and innovation program under the Marie Skłodowska-Curie grant agreement No. 860627 (CLARIFY Project).

The work of Adrián Colomer has been supported by the ValgrAI – Valencian Graduate School and Research Network for Artificial Intelligence \& Generalitat Valenciana and Universitat Politècnica de València (PAID-PD-22).

There are no material financial or non-financial interests to disclose for the authors.

\bibliography{refs} \bibliographystyle{unsrt}
\end{document}